\documentclass[sigconf,10pt]{acmart}
\usepackage[utf8]{inputenc}

\usepackage{color,colortbl}
\definecolor{Gray}{gray}{0.9}
\usepackage{balance}
\usepackage{url}

\title[COVID-19 Technology Concerns]{  2020 UK Lockdown Cyber Narratives: the Secure, the Insecure and the Worrying }
\author{Karen Renaud$^1$, Paul van Schaik$^2$, Alastair Irons$^3$, Sara Wilford$^4$}

\affiliation{$^1$Abertay University, Rhodes University $^2$Teeside University, $^3$Sunderland University, $^4$De Montfort University}
\email{cyber4humans@gmail.com}

\date{June 2020}

\begin{document}
\begin{abstract}
    On the 23rd March 2020, the UK entered a period of lockdown in the face of a deadly pandemic. While some were unable to work from home, many organisations were forced to move  their  activities online. 
    Here, we discuss the technologies they used, from a privacy and security perspective.  We also mention the communication failures that have exacerbated uncertainty and anxiety during the  crisis. 
    
An organisation could be driven to move their activities online by a range of disasters, of which a global pandemic is only one.      We seek, in this paper, to highlight the need for organisations to have contingency plans in place for this kind of eventuality. 
    
    The insecure usages and poor communications we highlight are a symptom of a lack of advance  pre-pandemic planning.  We hope that this paper will help organisations to plan more effectively for the future. 
\end{abstract}


\maketitle

\section{Introduction}
The pandemic of 2020 led countries to impose lockdowns, which closed schools, universities and a host of other workplaces.  This forced organisations to move their activities online, and employees were often left to find the best technologies to carry out their core activities. Many grasped at the most familiar or popular technologies to satisfy their needs. In some cases, the technologies they used, or the way they used them, exposed them to the actions of hackers, or violated their privacy. We explore these issues in Section \ref{streaming}.

On the other hand, the 2018 GDPR regulation made organisations consider and manage the way they stored  personal and sensitive data. 
The fact that organisations had gone through this process
stood them in good stead when the lockdown was imposed. This demonstrates the immense value of planning and putting  measures in place. 
We discuss  secure measures in Section \ref{secure}.


There were also some worrying developments, in terms of privacy and human rights violations. We discuss these in Section \ref{worrying}. 

\begin{figure}[ht]
    \centering
    {\includegraphics[width=0.9\columnwidth]{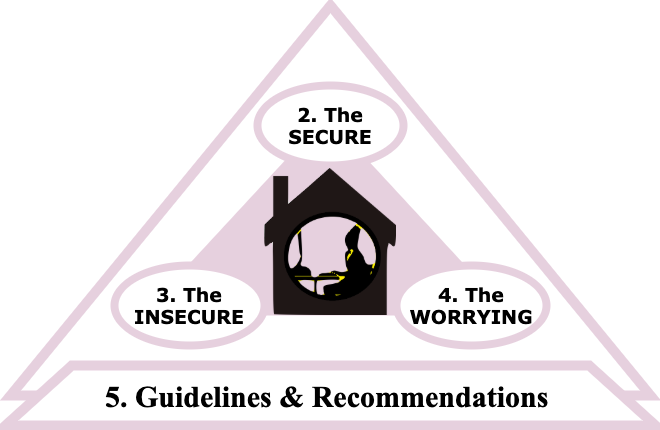}}
    \caption{Narratives Explored in this Paper}
    \label{fig:pandemic}
\end{figure}

Finally, in Section  \ref{discuss}, we present  guidelines for measures organisations could implement to prepare for future lockdowns. These will minimise the security and privacy violations that are occurring during the current lockdown, as revealed by our narratives. Section \ref{conc}
concludes. Figure \ref{fig:pandemic} provides an overview of the paper.

 The research methodology used in this paper is desk research. In other words, we gathered facts and news reports  that helped us to construct the emergent security and privacy focused narrative around the UK's 2020 pandemic lockdown. By so doing, we have been able to highlight the need for organisations to support their employees more effectively so that they can achieve private and secure at-home working. Bill Gates predicts that pandemics will occur every 20 years \cite{Crump}, and other disasters, such as fires, occur unpredictably and also send workers home.A lockdown could happen again.  We thus build on the research literature to propose  guidelines for working securely and privately at home, as our main contribution.  

\section{The Secure}\label{secure}
As academics, we are in a good position to write about the plans our Universities made before the introduction of GDPR in 2018. 
Universities across Europe engaged in activities to ensure that their employees knew exactly how to secure their data.

Our institutions implemented measures that are likely to be typical of the industry at large. 
They published policies and provided secure storage in some cases. Whatever their individual arrangements,  this preparation for GDPR meant they were well prepared for the lockdown, when it came to data storage. 

A brief review of the authors' institutions' policies  evidences this. 
One author's institution has a GDPR policy that lays down ``good practice'' principles for storing and managing research data. Guidance is also provided with respect to where data should be stored (i.e. on University sanctioned storage drives).
Another author's institution's information security policy devotes two pages to explaining how research data ought to be secured. A third author's institution has a five page
Data Protection Policy laying out guidelines for securing data. A fourth author's institution publishes an 9-page policy to guide data protection activities. 

 One policy specifies that OneDrive be used to store data, but the others do not do this. Two forbid the use of Dropbox for research data.
 
 As well as policies that address GDPR needs, there is also a need to consider the security of the architecture that enables home working. For example, considering Wifi routers: whether or not these are password-protected and the age of the routers is relevant (more recent routers have more inbuilt security). Moreover, others in the household may have access to the computers used for remote working  and the separation of work  and personal technology usage becomes challenging. 
 
 In none of the policies we reviewed was video conferencing software mentioned. Nor were any recommendations made about protocols for home working or secure software  to use, nor were there any guidelines for hardware configurations that would help improve home working security. Finally, there was no mention of the use of VPNs to preserve privacy.

\section{The Insecure}\label{streaming}
\subsection{Video Conferencing Technologies}\label{zoom}
The move to home working, and the use of online conferencing software, has increased exponentially \cite{Lenihan} as society embraces social distancing and heeds isolation instructions from the government. Quarantined employees naturally sought a way to stay in touch with their loved ones, and maintain relationships with colleagues, whilst also conducting their work activities from home.

\begin{figure}[ht]
    \centering
    \frame{\includegraphics[width=\columnwidth]{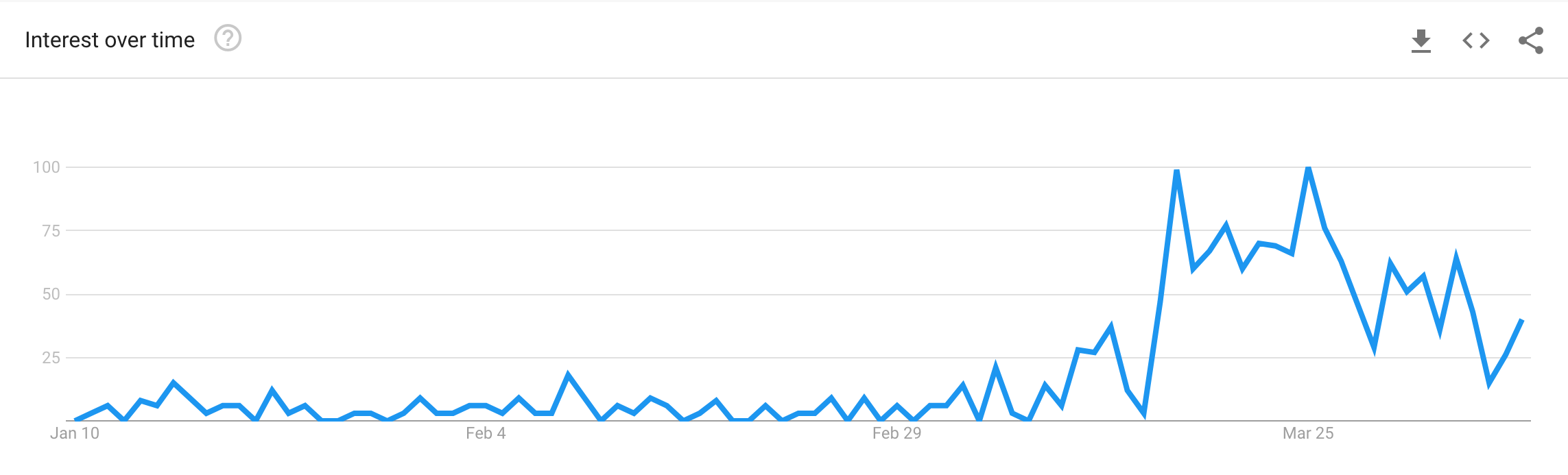}}
    \caption{Google Searches for ``Video Conferencing'' from 10 March 2020 to 10 April 2020}
    \label{fig:trends}

    \centering
    \frame{\includegraphics[width=\columnwidth]{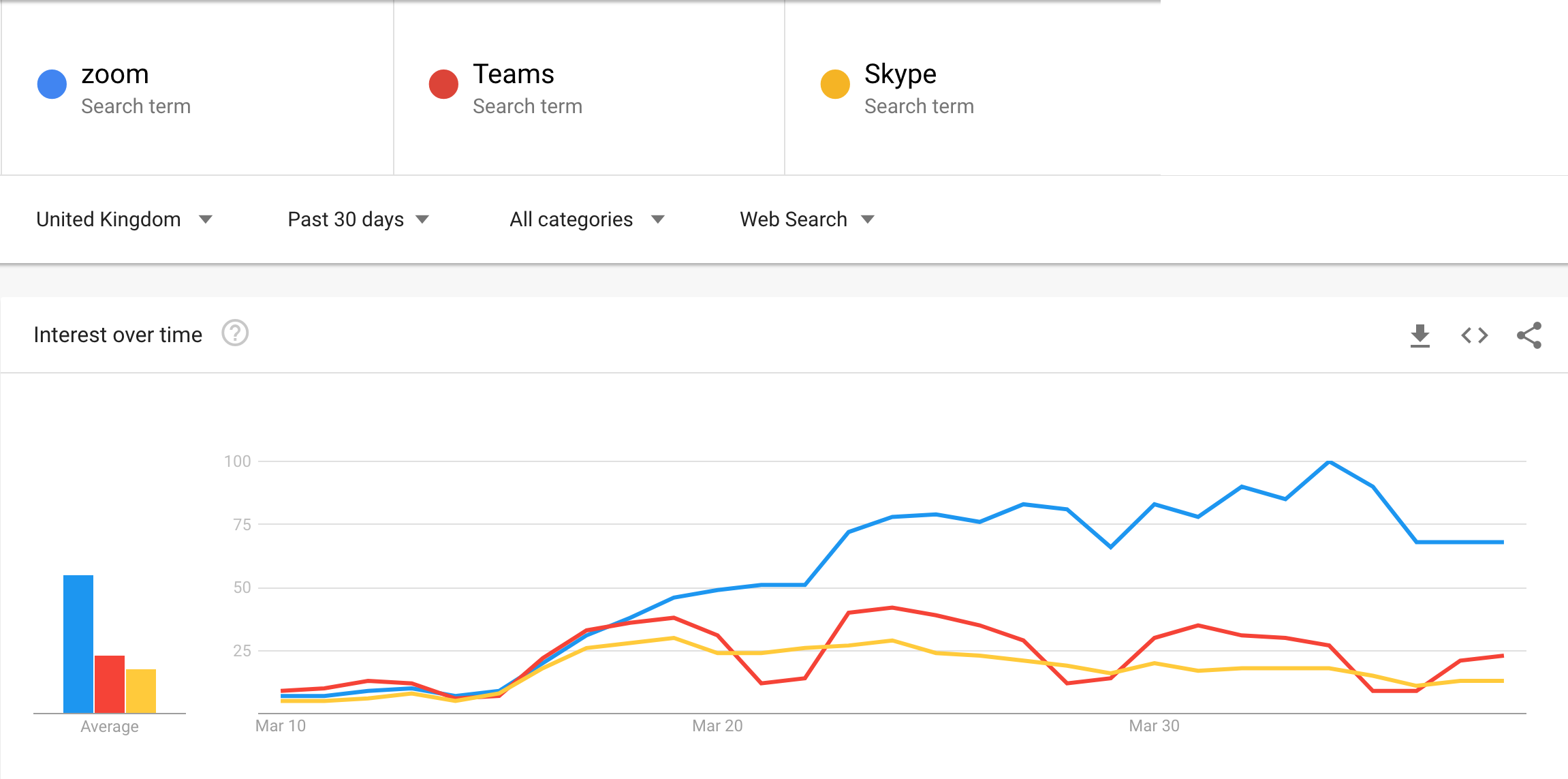}}
    \caption{Searches for Zoom, Teams and Skype from 10 March 2020 to 10 April 2020}
    \label{fig:zoom}
\end{figure}

There are a range of tools on that enable colleagues and collaborators to work together in a virtual ``face to face''  environment whilst maintaining social distancing. 
Many turned to Google to find the tools they needed. 
Figures \ref{fig:trends}  and \ref{fig:zoom} depict the spike in searches for video conferencing technologies since the beginning of the lockdown. As these graphs demonstrate, many found Zoom and looked for more information. 

The rush to adopt video conferencing technologies ushered in a number of interesting and concerning aspects, in terms of user behaviour.

\subsubsection{No Guidelines}\ \\
Because of the speed in implementing video conferencing technologies, very few organisations  had the time to put  policies or protocols in place regarding remote working practices. This means employees are figuring out their practices for themselves. They might be well informed, and do so securely, but they might also put themselves and their employers' devices at risk.  

\subsubsection{Unwise Installation}\ \\
 One the one hand, there is a need to be included and to ensure that users have the appropriate hardware to enable the software to be utilised.
 Users may have been given admin rights to enable them to install software because their corporate IT support do not have the capacity to support a multitude of users trying to install software. This may lead to advised checking and testing not being completed. It also potentially causes stress when things do not work or users cannot connect to the technology that their peers are using. 

On the other hand, the need to connect with others can lead to unwise installation of technologies and this may mean that the usual security checks are not in place. The adoption of workable solutions in the short term could well obscure underlying problems, for example with user privacy, data and information sharing, and possible unwitting GDPR violations. 

Furthermore, once the pandemic period is over, the extent of privacy, security and data protection violations will be discovered. The subsequent fall-out in terms of managing the administration of both criminal and civil regulations, could prove to be overwhelming to institutions, regulators and individuals, if the use of these emergency tools can be shown to have been used uncritically and without informed consent or sufficient care for personal data.

\subsubsection{Zoom}\ \\
This particular tool has become very popular. As its popularity has risen, so have concerns about its security and potential privacy violating practices \cite{Finnegan}. 
A number of security experts have raised concerns about vulnerabilities within Zoom
 \cite{Lakshmanan,Whittaker} 

$\bullet$ \emph{\textbf{Potential privacy violations}}. Research by Citizen Lab found that   \emph{cryptographic operations were delivered to participants via servers in China} \cite{Lee2020a,Lee2020b}.

$\bullet$ \emph{\textbf{``Zoomboming''}}.  This is where trolls take advantage of open or unprotected meetings and poor default configurations to  \emph{broadcast porn or other explicit material}. In response, Zoom  enabled the Waiting Room feature which allows a meeting host to control when a participant joins the meeting and enforces passwords.  On the 8th April researchers revealed a security vulnerability in the waiting room \cite{Marczak}.

$\bullet$ \emph{\textbf{Security vulnerabilities.}}  Researchers discovered a flaw in Zoom's Windows application which  \emph{allowed 
remote attackers to steal victims' Windows login credentials} and  execute arbitrary commands on their systems. A patch was issued on April 2, 2020, to address this flaw.
Other researchers created a new tool called ``zWarDial'' that searches for open Zoom meeting IDs,  \emph{finding around 100 meetings per hour that aren't protected by any password}. There is also evidence that Zoom IDs and passwords are being sold on the dark web \cite{Binder}.

As we write, some countries  \cite{Sephton} have suspended the use of  Zoom by teachers due to abuse by hackers. 
Zoom has been particularly responsive to these criticisms, patching them as quickly as they can \cite{Mudge}. 
However, as Mudge \cite{Mudge} points out, whilst the Zoom developers should have addressed security issues in the initial design, 5 years ago, they are trying to ``fix'' vulnerabilities as the Zoom user community grows. Mudge also indicates that there is a responsibility on the user to ensure security of any application that they are using, including Zoom. Making sure that routers are robust (and up to date), making sure tools are up to date and ensuring the patches and updates are put in place as soon as they become available will all contribute to safer Video Conferencing environments. 
The Zoom developers will also see this as an opportunity to improve their product. 

The Zoom scenario seems to be a classic trade off between usability and security, highlighted by Cranor and Garfinkel \cite{cranor2005security}. The rush to ensure that colleagues could stay connected, at very short notice meant that easy to install and  use applications, such as Zoom, are being used without people thinking about security or even knowing what steps to take to assure secure usage.

\subsubsection{In Conclusion}\ \\
We have discussed Zoom, which has attracted a great deal of attention due to its escalating user numbers. It is likely that many of the other video conferencing offerings also have vulnerabilities which are, even now, being exploited by hackers. 

Our argument is not Zoom-specific. We are making the point that many  are using a variety of video conferencing technologies with serious vulnerabilities, and they are doing so because they either do not have any alternatives, because they feel impelled by the critical mass usage to use them, or because they are simply unaware of the vulnerabilities. 

\subsection{School Teachers}\label{teachers}
School teachers are perhaps most unprepared to move their activities online. The bulk of their work is face to face with the children, in and out of classrooms. Now suddenly they have had to find other ways to engage. There is some evidence that they are woefully uninformed about privacy \cite{Gym} and security \cite{pusey2011cyberethics}. 

The advice from Human Rights Watch \cite{Martinez} is to focus on the most accessible technologies and methods. There is no mention of privacy or security considerations in this article. Yet privacy, too, is a human right (United Nations Declaration of Human Rights (UDHR) 1948, Article
12) and in Europe the new GDPR regulations have stringent rules requiring that children's data be kept private \cite{GDPR}. We  now provide two examples of teachers maximising accessibility with security and privacy pitfalls.

 \subsubsection{Example: Gym Teachers}\ \\
Gym teachers would like their pupils to provide evidence that they are doing their exercises. How do they do this while all their pupils are quarantined?

 Anonymous \cite{Gym} 
 posted a comment to a Reddit group, saying that a teacher wanted children to post videos to YouTube.  
 
 The first comment is from a teacher, defending the practice: ``\emph{I think the mindset is trying to prevent students from just faking it by having them show evidence, but an email to the teacher asking for an alternative assignment should work.}'' Another teacher explains: ``\emph{My district wants me to document everything I am doing daily that is focused on my degree. 3 to 5 hours a day M-F. How the hell can I come up with 3 to 5 hrs of stuff 5 days a week?! We have to document it and turn in a log sheet every Monday.}'' 
 
 One commenter doesn't understand what the fuss is about: ``\emph{Maybe not social media but why not? Meet them where they are at and remove barriers.}''
 
   Some  suggest alternatives: 
  ``\emph{I friend of mine made a private Facebook group for his classes and that’s how they’re going everything}.'' However, Facebook and privacy are diametrically opposed \cite{Facebook}. 
  
  Another offered advice: 
  ``\emph{YouTube videos can be private - accessed only with a link. If this must happen, that could be an option - then delete the video after the grade is marked.}'' 
  The question that has to be asked, in this case, is how the password will be communicated to the teacher? If email is used, the YouTube video is not private. 
  
 Others express more concerns about privacy violations:
  ``\emph{Nothing is truly private on the internet and there are a bunch of bored perverts home with nothing else to do}.'' 
  
 The upshot is that
children's privacy is being lost.\\
 \ \\
 \textbf{\emph{Don't Blame the Teachers}}\ \\
  One of the final comments expresses the unreadiness of this demographic. 
  ``\emph{Teachers are working their asses off to put a plan in motion that hasn’t even been finished yet. The nation, state and district haven’t planned ahead for this, it’s all being crafted on the spot and teachers are the ones doing the brunt of the work to make sure your nephew, niece, child, loved one is still being educated, still given a routine, still knows they have someone checking in on them and supporting them. Get off your reddit high horse and thank a [snip] teacher who has been sourcing every ounce of ``creativity'' to meet the demands of the education system with minimal support and with educ.}''
  Certainly, many parents are expressing their support for the huge efforts teachers are making \cite{Deabler}. 
  
  We are not blaming teachers; we are pointing the finger at those who are responsible for providing teachers with the technologies they need to carry out their activities during the pandemic. Based on the evidence we have gathered, it certainly seems as if they have been left to find their own way. That they make mistakes when their employers fail to support them is understandable.
  
 \subsubsection{{Example:  Higher Education}}\ \\
  Higher education institutions have also been forced to provide teaching, resources and support activities online. The concerns about privacy, data protection, verification, collusion, cheating, and how to conduct online assessment, are similar to those experienced by school teachers. Higher education institutions however, have access to vastly more resources than schools, including an existing distance learning infrastructure, and a relatively high proportion of staff able to work from home. 
  
  In responding to the crisis, Universities offered access to multiple online tools, often without due diligence of privacy and security issues, or guidance for staff. There did not seem to be time to think about this in the rush to go online, with the focus being on delivery, staff/student support and promoting key public health advice \cite{UniversitiesUK2020}. 
  
  Whilst this is understandable, the lack of oversight may result in significant problems in the future, as video conferencing apps, and an eclectic mix of online tools are used without a second thought (the Zoom app is integrated into Microsoft teams \cite{Microsoft2020}). The result is that confidential discussions are open to interception, student work may be accessed, exams compromised and sensitive data inadvertently leaked. 
  
  However, there is some awareness of the issues, and academic institutions worldwide are working to understand and make sense of this new way of working \cite{Lim2020}. Meanwhile, the technology companies are  under pressure to address concerns and to secure their systems \cite{Hern}, but it is likely to be too little, too late.

\section{The Worrying}\label{worrying}
\subsection{Contact-tracing Apps}\label{contacttracing}

Governments are doing everything they can to prevent deaths during the COVID-19 pandemic. Some governments have used contact tracing tools, either by mobile phone apps or using cell tower triangulation \cite{Korea,Singapore,Volpicelli,Kelion}. These can trace all the people an infected person has been in contact with to provide warning of possible infection. Contact tracing is a mature technique, which has been used to track tuberculosis  \cite{TB}, SARS \cite{ferretti2020quantifying} or STDs \cite{bell2011partner} contacts. 

Yet these apps, even if initially justified during the pandemic, can very quickly violate privacy and other human rights after the pandemic has abated \cite{CANCRYN}. 
Some countries have used contact tracing in an authoritarian and privacy violating fashion \cite{Zastrow,Brookes2020,Landau}. 

An Israeli company has published an app \cite{Cluley} which claims to respect the privacy of citizens, as follows (direct quote from Cluley):

$\bullet$ 
Use of the app is optional, not compulsory.
Any location data collected by the app does not leave the phone, and is not uploaded to the Israeli government. All processing happens on the phone itself.

$\bullet$ Those diagnosed with Coronavirus have to volunteer their location history for use by the app, which is driven by a JSON file that is updated with new data on an hourly basis.

$\bullet$  Even if a match is made, the app does not inform the Israeli Ministry of Health. It’s up to the user to get in touch if the app alerts that there might have been an encounter with a Coronavirus case.

$\bullet$  To reassure users about the behaviour of the app, it has been released as open source and its code published on Github.

$\bullet$  The app’s code has been examined by security experts at Profero.


All of these efforts, and the motivation to use an app, are based on the assumption that infection can only occur within 6 feet of an infected person. Yet MIT recently published research that showed that the droplets from a cough or sneeze could travel up to 27 feet \cite{Culver}.

The immediate justifications for extending surveillance of the public, uses the rhetoric of war to `battle the virus' and to reinforce its citizen's sense that `we are all in this together' \cite{sun2016war}.
but the expectation that all good citizens should be happy to utilise the app, and therefore give up some of their liberties for the benefit of society \cite{etzioni2004communitarian}.
This then helps to create a moral imperative to comply with requests for quite draconian restrictions on civil liberties. By using `nudge' tactics \cite{mulderrig2018multimodal}, they attempt to habituate the population into acceptance of  increased electronic surveillance. This  means that those not engaging can  be presented as having a moral failing or a lack of civic awareness. 
\begin{figure}[b]
    \centering
    \frame{\includegraphics[width=\columnwidth]{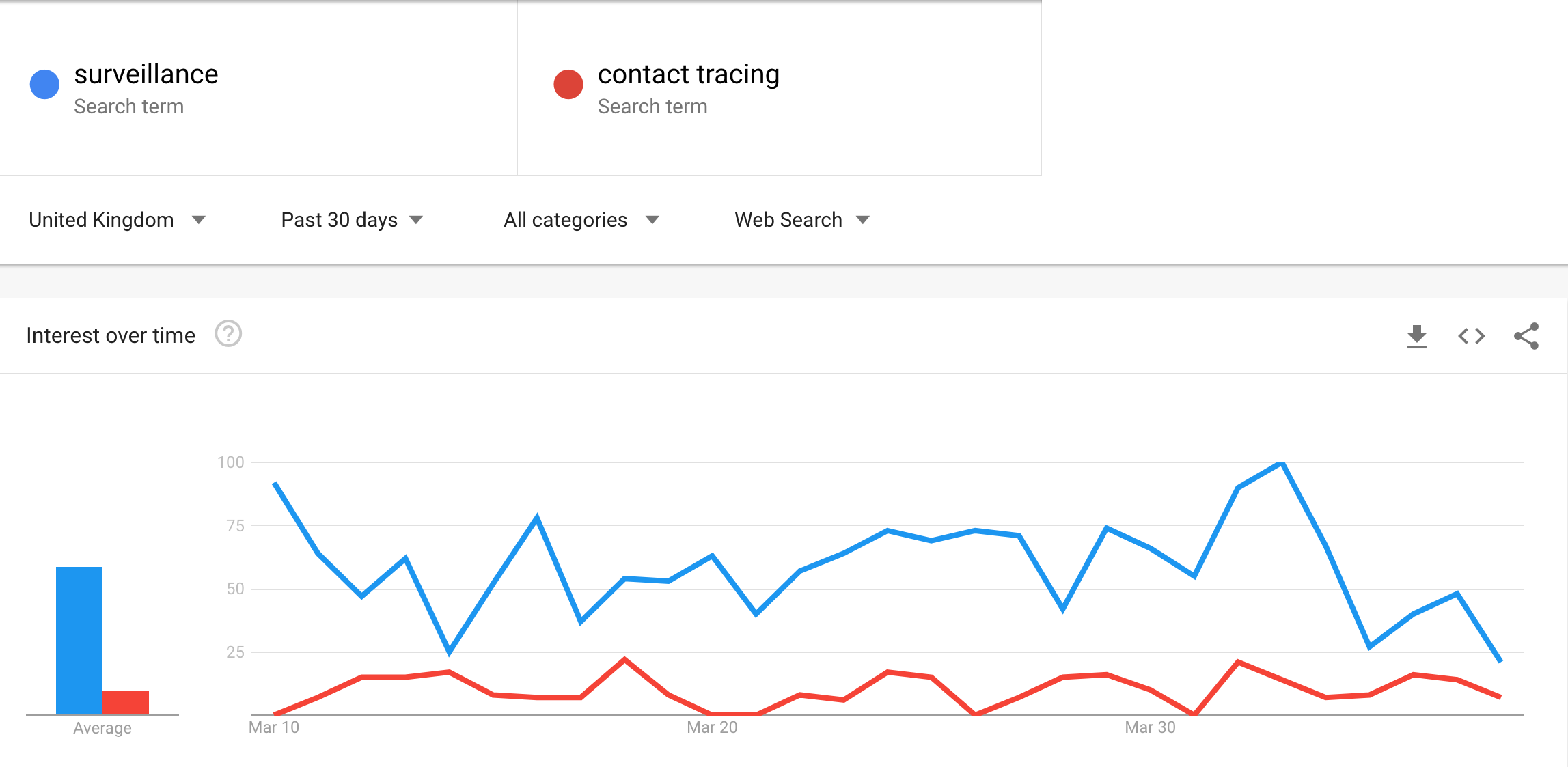}}
    \caption{Google Searches for Surveillance vs. Contact Tracing from 10 March 2020 to 10 April 2020}
    \label{fig:surveillance}
\end{figure}
In terms of the `battle' against COVID-19, the expectation is that everyone will do their utmost to prevent its spread, and dire warnings, accompanied by daily death rates, will serve to sustain pressure to submit to increased surveillance. Social shaming regarding the lock-down conditions is already evident, encouraged and actively promoted by the media \cite{Hawken-2020}. Social media is further increasing the pressure, and includes trolling, and other forms of online abuse, aimed at those perceived to question government requests. 

This approach is not dissimilar to an oft-repeated statement used to  shut down concerns about the use of surveillance, particularly post 9/11: ``\emph{if you’ve got nothing to hide, you’ve got nothing to fear}'' \cite{solove2011nothing}. The message here is that surveillance that enhances national security and protects against terrorism should take precedence over personal liberty. It can be surmised that it is only a matter of time before such rhetoric is resurgent in the public dialogue. In a crisis, most people are eager to help their fellow citizens \cite{Ward} and  compliance with rules is consistent with this stance. However,  a few dissenting voices are expressing concerns about the potential future use of these technologies  \cite{Myers}. 

The acceptance of increased surveillance may not be encouraged just through the impact of a global emergency, social shaming and nudge tactics. In many countries, both authoritarian and democratic, people have become habituated to living under surveillance via CCTV, GPS, Smart Phones and during Internet usage, whether that is by the government \cite{dinev2008internet,walton2001china} or by businesses \cite{zuboff2019age}. Many  have become so accustomed to being under surveillance, that the Panopticon, a prison system of total surveillance and a `new mode of obtaining power of mind over mind' devised by Jeremy Bentham  
 \cite{bentham1995panopticon} has become the reality of our modern, surveillance society \cite{foucoult1975discipline,lyon2017bentham}.
 
 As the current situation begins to resolve,  questions about an end to the increased surveillance will be raised. It is likely that arguments will then be made to retain these technologies in the fight against crime or terrorism and above all, to ensure that ``\emph{We are ready next time}". In response, parliaments have an important role to hold the government to account.  The UK Labour opposition  leader has stated that he will scrutinise the UK government's actions and point out any failures that need to be addressed \cite{Starmer2020}. However, the key to the success or failure of current and future responses to such a crisis lies in how and what information is communicated to the public.

\subsection{Ineffective Communication}\label{poorcomms}
\subsubsection{Government Communication to the Police} \label{police}\ \\
The UK government passed coronavirus legislation, which gave police new powers \cite{NewLaws}. Very soon after the country was put into lockdown, reports of police men and women exceeding their remit began to emerge. 


For example, Derbyshire Police used drones to film people walking in the hills, on their own, to name and shame them online \cite{DerbysPolice} and dyed the local pool black so that people would not want to take a swim \cite{DerbysPolice2}. This police force is not the only one to overstep the mark. 

Warrington police posted to Twitter that `six people had been summonsed for offences relating to the new coronavirus legislation to protect the public’ \cite{Warrington}. The violations include
going `\emph{out for a drive due to boredom}’ and `\emph{multiple people from the same household going to the shops for non-essential items}'. The legislation does not specify what essential items are, so the police are clearly deciding for themselves. For example, a news report on the 30th March reported that police had ruled Easter eggs ``non-essential'' \cite{Ellson} (in the week before Easter). Wine and crisps too, were ruled non-essential \cite{Mitchell}. Given that the government requires that citizens do not do non-essential shopping, would it be the shops that are responsible to ensure that non-essential items are not offered for sale?

Slater \cite{Slater} argues that: ``\emph{The thing is when you give police – or in the case of these new regulations, police, community support officers and other people ‘designated’ by local authorities – the power and moral authority to throw their weight around, many of them are bound to overinterpret their responsibilities and overstep the mark. }'' This sentiment is echoed by Campbell \cite{Campbell}.

Indeed, police chiefs have now become concerned enough to issue a statement saying they will be drawing up new guidelines so that their police forces \cite{Milne} do not overreach their authority. This was in response to former Supreme Court Justice Lord Sumption saying that Britain risked turning into a ``police state'' \cite{Carroll}. 
The Home Secretary Priti Patel was moved by comments issued by Northamptonshire Police Chief Constable Nick Adderley to issue a warning to police on the 10th April saying that  road blocks and checking of supermarket trolleys were ``not appropriate'' \cite{Speare-Cole}.

We believe many of these issues can be traced back to poor communication from the UK government about what actions the police should be taking. In the absence of clarity, some people will naturally overreach, as indeed they have. 

\subsubsection{Government Communication to Citizens}\label{govt}\ \\
Grater \cite{Grater} reports on Emily Maitlis calling the UK government's language `trite' and `misleading', when they were discussing  the COVID-19 virus.
She said that  UK Cabinet member Dominic Raab erroneously suggested that 
people could survive the illness through fortitude and strength of character.
She also pointed out that the virus and the lockdown was much harder on the poor, than on the wealthy. 

During the week of the 6th April 2020, every household in the United Kingdom received a letter from the Prime Minister.

\begin{figure}[t]
    \centering
    \includegraphics[width=\columnwidth]{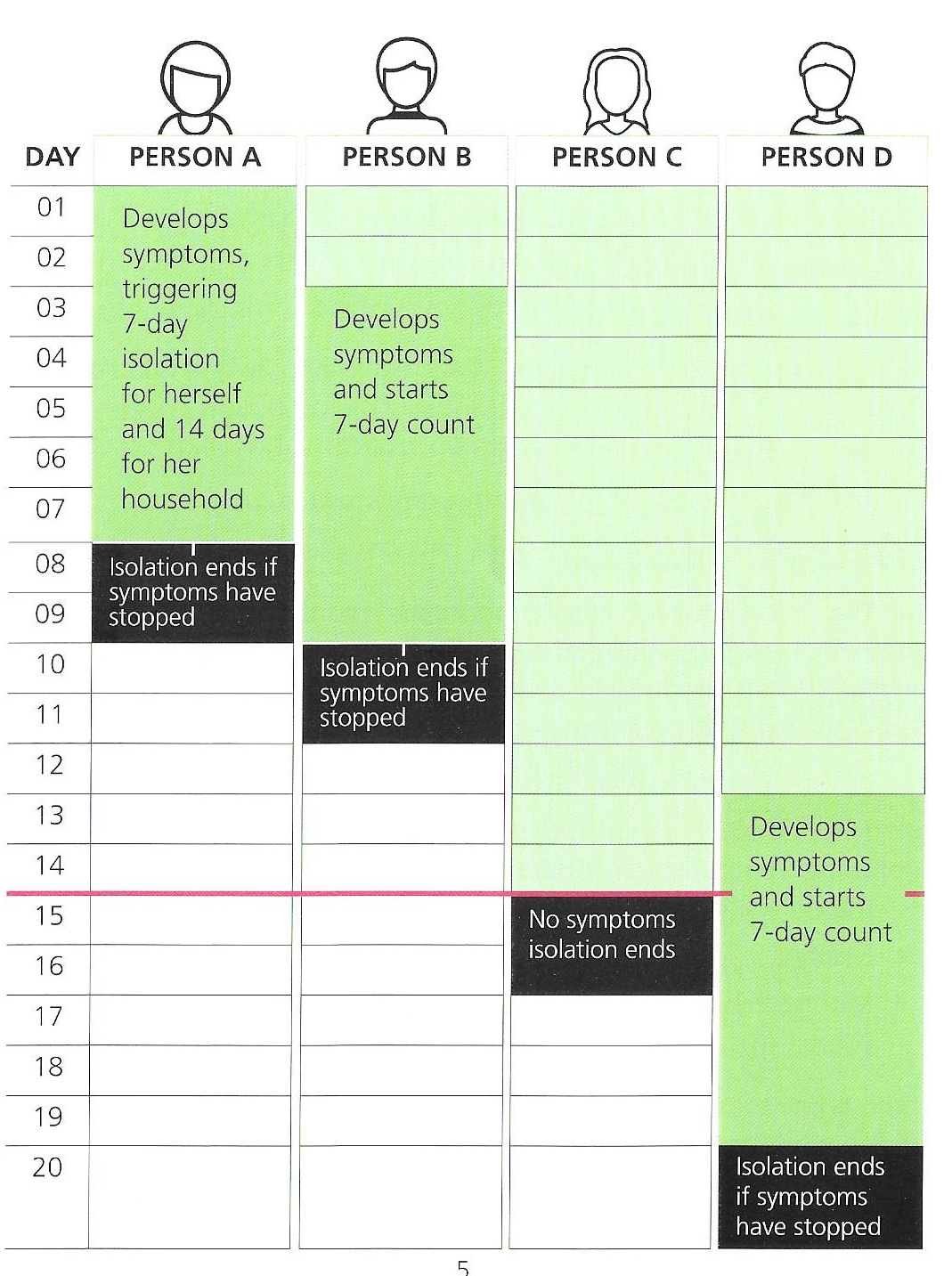}
    \caption{CORONAVIRUS Stay at Home Protect the NHS Save Lives}
    \label{fig:graph}
    
        \centering
    \frame{\includegraphics[width=\columnwidth]{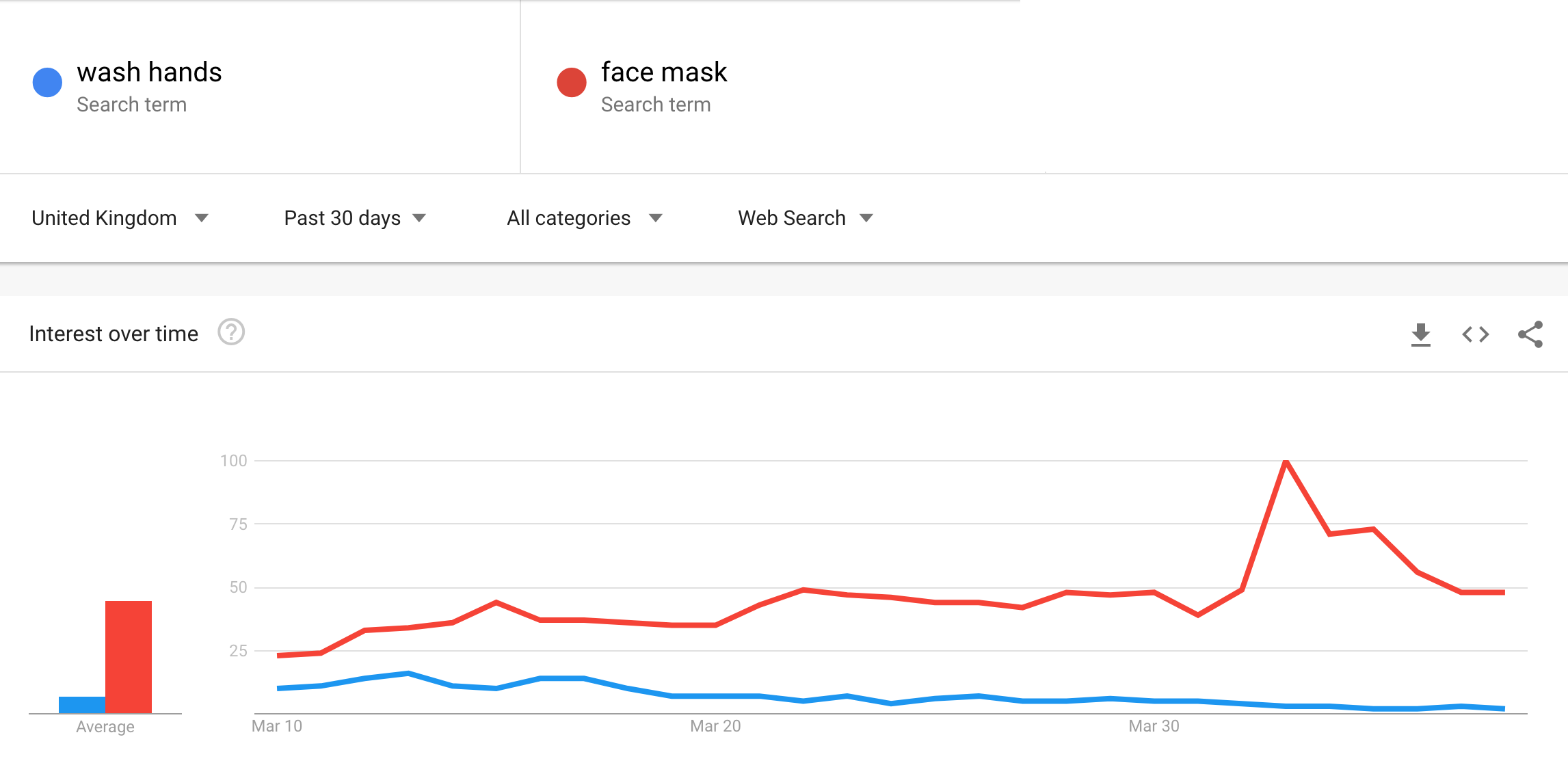}}
    \caption{Hand Washing vs. Face Masks}
    \label{fig:masks}
\end{figure}

With the letter was a leaflet which included the diagram in Figure \ref{fig:graph}. This graph was rather puzzling and seemed inaccurate since the letter came from the Prime Minister, who himself had just
been admitted to hospital \textbf{10 days} after falling ill with the virus. 
This diagram conflicts with the text in the leaflet, which suggests that Person C should isolate for an extra 7 days, now that Person D has started exhibiting symptoms. 

 From the very beginning, when the coronavirus emerged in Wuhan, the one consistent message has been to wash hands. On the other hand,  the communication around face masks has been confusing and inconsistent. The Google Trends graph in Figure \ref{fig:masks} demonstrates the uncertainty around face masks, as evidenced by increased numbers of Google searches. 
People were initially told that face masks were ineffective \cite{Jones}. Then, on the 1st April, the World Health Organisation (WHO) announced that it was considering changing its guidance on face masks \cite{Devlin}.
On the 7th April, the Centre for Disease Control (CDC) recommended wearing face masks in public \cite{Dolcourt}.
On 8th April, the WHO announced that there was no evidence to suggest wearing a face mask would prevent healthy people from catching Covid-19 \cite{Gallagher}.
These kinds of conflicting messages have led to a great deal of uncertainty.

\section{Guidelines \& Recommendations}\label{discuss}
Organisations should have contingency plans to support home working. This ought to include  a suite of technologies that people are given to ensure that their online activities are carried out securely. Moreover, these ought to be tested to ensure that employee privacy is preserved. If people are under pressure to do their jobs, without being given the necessary tools, they are likely to find a way. 
Humans are endlessly innovative in overcoming obstacles and finding a way to fulfil their commitments. 

Here, we present some guidelines and recommendations regarding security and privacy for private companies, public organisations and third-sector organisations in response to the COVID-19 crisis.

\subsection{Sustaining the Secure}
 Organisations should build on their existing  information security and privacy policies by updating these to ensure they cover unforeseen problems that have now been exposed by the pandemic lockdown.  Issues to be addressed include video-conferencing and other tools used to support remote working, teaching or learning from home  and the management of signing up to applications or websites. 

As it is unlikely that staff will fully read and/or understand the complete policy, organisations should consider creating a summary of topics that are specifically relevant to the crisis, with clear advice, and communicate these in a straightforward way to their staff, with a link to full details within the applicable policy.  

Based on existing research evidence \cite{harris2008living}, this could be an effective way of positively influencing staff behaviour to reduce security and privacy problems.   Similarly, governments should build on existing guidance by updating this for their citizens in terms of information security and privacy, and communicate the main points in a straightforward manner, with a link to full details. This will alleviate the problems highlighted in Section \ref{govt}.

\subsection{Securing the Insecure}
\textbf{\emph{Technology Choice:}} A co-ordinated approach is needed to analyse and identify the security and privacy strengths and weaknesses of video-conferencing  and other networked tools (Section \ref{streaming}), and then to publicise the results with straightforward guidance to the general public. 
This is important to ensure that users use appropriately secure and privacy-protecting tools and appropriate tool configurations to protect their information  and preserve their privacy.

\textbf{\emph{WiFi:}} Consider the use of WiFi.   Employees should avoid the use of public WiFi and make sure that home WiFi is as secure as possible. Attempts should be made to ensure that routers in the home are password-protected and, where possible, those working at home should have as up to date a router as possible. More recent routers, those less than 5 years old, have more built-in security. 

\textbf{\emph{Authentication:}} Those working from home should follow the advice given by their organisations’ IT services and use approved (and supported) software. It is good practice to make use of strong passwords and ensure that, where possible, multi-factor authentication is utilised. Make sure  there is up to date antivirus software in place and install updates and patches as soon as they become available, as long as these are from official sources. 

\textbf{\emph{Access Control:}} In addition, every attempt should be made to separate work and personal devices and people should be advised not to share work technologies with family members.  



In this co-ordinated approach, there may be roles for third-sector organisations that specialise in information security and/or privacy and for government security organisations.  It is important that the communication of the guidance be clear and available from a single access point.  This will allow users to more easily locate, assimilate and apply the guidance \cite{valenti2018being}.

\subsection{Alleviating the Worrying}
\subsubsection{Privacy \& Human Rights}\ \\
The development of a wide variety of contact-tracing apps (Section \ref{contacttracing}) could be useful in terms of promoting competition and innovation to improve the quality of app support for combatting the virus.  However, given the crisis at this time, collaboration can be more beneficial than competition \cite{Boss2019}, as the latter:

(1) may be too slow to improve quality on time, as the virus does not follow the competition’s timetable,

(2) may prioritise the app’s ‘virus-combat effectiveness’ at the cost of information security, privacy and other considerations, and 

(3) may be too slow in terms of take-up for people to choose the best possible app because of information overload (too many apps on offer with too many choice attributes to consider) \cite{lee2016information} and choice inertia (people are slow or reluctant to change to another product even if it is better) \cite{ashby2019effect}.

Therefore, in crisis times such as this, instead a collaborative multidisciplinary approach should be followed to ensure that effectiveness, security, privacy, usability and other considerations are taking into account.  

In addition, iterative design and testing is necessary to ensure quality improves during development, but also after deployment to continually improve the app \cite{cranor2017putting,emami2020ask}.  

A single app per country or coalition of countries (for example, the EU) would provide a single access point.  This will allow users to more easily locate, assimilate and apply the information.  An example of a pan-European approach to develop a GDPR-compliant app is currently in operation under the name Pan-European Privacy-Preserving Proximity Tracing (PEPP-PT2020){\footnote{\url{https://www.heise.de/newsticker/meldung/Corona-Tracking-Apps-mit-PEPP-PT-Entscheidend-ist-fuer-uns-dass-der-Datenschutz-gewaehrleistet-wird-4700336.html}}}.

More generally, from the perspective of privacy, several evaluation criteria have been identified that specifically contact-tracing apps should meet \cite{Montjoye2020}. These are:
$\bullet$ 
limiting the personal data gathered by the authority;

$\bullet$ 
protecting the anonymity of every user;

$\bullet$ 
    not revealing to the authority the identity of users who are at risk;
preventing the system can be used by users to learn who is infected or at risk, even in their social circle;

$\bullet$ 
preventing users from learning any personal information about other users;

$\bullet$ 
preventing external parties from exploiting the system to track users or infer whether they are infected;

$\bullet$ 
putting in place additional measures to protect the personal data of infected and at risk users

$\bullet$ 
providing support for people to verify that the system does what it says.

Contact-tracing apps will vary in the extent to which they support privacy.  For example, the Pan-European Privacy-Preserving Proximity Tracing (PEPP-PT) app \cite{PEPP-PT2020} aims to preserve privacy and maintain security, but does not explicitly address all these criteria.

\subsubsection{Communication}\ \\
Regarding communication weaknesses (Section \ref{govt}), any measures and the communication of these measures need  first to be thought through thoroughly and they should be based on the nationally and/or internationally available relevant expertise.  Therefore, the recommendation is to make good use of academic and other experts in relation to the content and communication of measures.  

In addition, the communication and measures should be thoroughly evaluated before they are released within the time constraints.  This is because poor communication can have adverse consequences in terms of potentially lengthening the pandemic, as our analysis of government communication shows.  It should be possible to do the required evaluation relatively quickly by drafting in academic and other experts to help combat the crisis. Whether their contribution is made mandatory or not, many of them will be keen to help contribute to solving the crisis, in any case, when given the opportunity.


\section{Conclusion}\label{conc}

These are unprecedented times and the move to home working has been implemented in a necessarily speedy way. However, as the initial implementation settles and people begin to get used to the ``new norm'', there is a need to reflect on the efficacy and viability of thereof. The stresses of  needing to participate in organisational activities mean that the ability to engage could lead to usual security measures and non-compliance with organisational policies.  Cyber criminals  will likely exploit  weaknesses and vulnerabilities in the home working environment. 

As has been illustrated by our narratives, the vulnerabilities can result from the lack of specific policies, consequent ``needs must work-arounds'', insecure hardware and/or non-robust software applications. These, combined with an increase in malicious attacks, mean that the home working environment is replete with potential digital threats.    

We do not seek to criticise anyone for the technologies they make use of in an emergency. We wrote this paper to highlight the difficulties employees faced and the security risks quarantined citizens unwittingly exposed themselves to. We  suggest a better way forward as the main contribution of this paper. 

\balance
\bibliographystyle{ACM-Reference-Format}
\bibliography{refs}

\end{document}